\documentclass[aps,pre,twocolumn]{revtex4-1}
\usepackage[T1]{fontenc}
\usepackage[utf8]{inputenc}
\usepackage{amsmath,amssymb,amsfonts}
\usepackage{graphicx,ifpdf}
\usepackage{relsize}

\newcommand\vB{\boldsymbol{B}}
 
\newcommand{\ud}[2][\!]{\,\mathrm{d}\sp{#1}{#2}\,}
\newcommand{\dirac}[2][\!]{\delta\sp{#1}\!\left(#2\right)}
\newcommand\e{\mathrm{e}}
\newcommand\ii{\mathrm{i}}

\newcommand\pP{\mathsf{P}}
\newcommand\vq{\mathbf{q}}
\newcommand\vP{\boldsymbol{P}}
\newcommand\pQ{\mathsf{Q}}

\newcommand\vR{\boldsymbol{R}}
\newcommand\R{\mathbf{R}}
\newcommand\vS{\mathbf{S}}

\newcommand\vx{\boldsymbol{x}}
\newcommand\vX{\boldsymbol{X}}
\newcommand\vy{\boldsymbol{y}}

\newcommand\derp[3][]{\frac{\partial\sp{#1}{#3}}{\partial{#2}\sp{#1}}}

\newcommand\moy[1]{\left\langle{#1}\right\rangle}

\DeclareMathOperator\erfc{erfc}

\begin{document}
\title{Local time in diffusive media and applications to imaging}
\author{Vincent Rossetto}
\email[Author's e-mail:]{\tt vincent.rossetto@grenoble.cnrs.fr}
\affiliation{Université de Grenoble I - LPMMC / CNRS,
\\25 avenue des Martyrs, 38042 Grenoble CEDEX 09, France}
\date{\today}

\begin{abstract}
Local time is the measure of how much time a random walk has visited
a given position. 
In multiple scattering media, where waves are diffuse,
local time measures the sensitivity of the waves to the local medium's
properties.
Local variations of absorption, velocity and scattering between two
measurements yield variations in the wave field. These variations
are proportionnal to the local time of the volume where the change happened
and the amplitude of variation. 
The wave field variations are measured using correlations
and can be used as input 
in a inversion algorithm to produce variation maps. 
The present article gives the expression of the local time in dimensions
one, two and three and an expression of its fluctuations, in order to
perform such inversions and estimate their accuracy. 
\end{abstract}
\maketitle

\section{Introduction}
Standard imaging techniques in physical sciences are based on the deterministic
nature of wave transport. In transmission, absorption 
or reflection imaging such as in the different types of optical microscopy,
radiography, ultrasonography, the production of an image relies on the fact 
that the source wavelength is much larger than the irrelevant and unpredictable
variations of the properties of the propagation medium. 
As long as only a small fraction of the energy is scattered by these 
irregularities, scattering is
negligible and the medium is \emph{transparent}. 
Media where scattering is not negligible
are called \emph{scattering media}. In scattering media a
substancial amount of the wave energy is scattered and more advanced imaging
techniques are necessary. Several strategies have been designed to obtain
accurate and reliable results. In optical coherence tomography
for instance, the fraction of signal that has been scattered more than once is
filtered out. However, when scattering is so strong that the largest part of
the signal has been multiply scattered, these strategies fail: the medium
becomes opaque.

In strongly scattering media, wave propagation reaches rapidly a
\emph{diffuse} regime: the intensity in space and time evolves
according to a diffusion equation. Diffusive media, such as fog, are commonly
believed to yield fuzzy images. However, it was recently realized that 
even in the diffuse regime, the measured signals carry informations. 
The wave recorded at a given position is the superposition of partial waves 
which trajectories follow a certain statistics determined by the diffusion
equation. These partial waves statistically explore the whole medium
and are sensitive to its local properties. 
The measured signal therefore contains informations concerning
the transport properties of the medium collected during their propagation, 
but these informations are
entangled in such a way that the raw experimental measurements cannot
be directly used to produce a regular image of the medium.

In heterogeneous, turbid media, techniques based on the correlations 
have proved efficient to extract some relevant informations. 
The nowadays commonly used diffusing-wave spectroscopy (DWS) 
allows one to measure the diffusion of scatterers~\cite{pine1988}
in diffuse regime for the waves.
However, these techniques remain poorly sensitive for imaging a
medium with static scatterers. It is nonetheless possible to 
increase the sensitivity if one considers the field rather than
the intensity~\cite{cowan2002}. The field is the displacement,
the pressure or the electric field, depending on the nature of the wave.
It can be recorded for waves with frequencies below the
order of THz, which are encountered  in acoustics, seismology or
material sciences. Modern recording devices provide precise
amplitude records for them at an appropriate, higher sampling frequency. 
The wave field records are used to compare two states of the same system at
different times and produce an image of the changes in the system,
this is why one speaks of \emph{differential imaging}.
The achieved sensitivity is so strong that a single change on one scatterer
among thousands of others can be detected.
Small relative velocity changes, of the order of~$10^{-5}$, have also 
been measured using a technique based on the diffuse part of acoustic records
\cite{larose2008,niederleithinger2013} or the late part of seismograms, 
the seismic \textit{coda}~\cite{pacheco2005,battaglia2012}.
These results have been obtained by measuring the time shifts between waveforms 
that maximize their correlations. The name of coda wave interferometry (CWI)
is often met to gather these approaches. 
It was recently shown that the maximum of correlation obtained in this way 
contains informations about the changes in the scattering properties
of the medium~\cite{rossetto2011}. The data extracted by correlation
are used as input in inversion techniques to produce images. 
Defects appearing in a concrete structure have been located correctly
by an error minimization inversion 
\cite{larose2010}. Local time has been used as a kernel in seismology
to observe the depth sensitivity of the seismic coda to 
velocity perturbations\cite{obermann}.
In a volcano, precursor structural changes to an eruption
have been observed using LOCADIFF~\cite{obermann}.

The underlying common principle of CWI and LOCADIFF is based
on the observation that the sensitivity of the diffuse waves to a change
in the medium is proportionnal to the time spent at the position
where the change has occured. This time is called the~\emph{local time}. 
Local time is a way to focus on a position in a heterogeneous medium. It is a
random quantity, because of the random nature of the wave propagation
in hereogeneous media, but its average value is well defined. However
the distribution of local time is defined only in one dimension.
To sort out the statistical error in practical applications, one needs 
to know the order of magnitude of the fluctuations. In the present work, 
I introduce the notion of local time and compute its average value in 
diffusive media. To obtain the fluctuations, it is necessary to introduce
a \emph{resolution length} on which the fluctuations depend.

The present article aims at introducing the concept of local time and at
giving the possibility to investigate its relevance in every field
of Physics where waves are strongly scattered.
Parts of this article are technical; they are intended to give a
complete material on the subject of local time and its fluctuations.
These parts can be skipped on the first reading and 
are pointed out along the text.
The article starts
with the introduction of the local time and some of its general properties.
A brief mathematical definition is sketched at the end of Section~\ref{general}.
The Brownian bridge is introduced elementarily in paragraph~\ref{b1}
and more mathematically in paragraph~\ref{GRqu}. 
Section~\ref{sec.average} briefly presents the computation
of the average local time in dimensions one, two and three using the
Brownian bridge properties. Only the paragraph~\ref{maths1}
is technical, the other paragraph of Section~\ref{sec.average}
being dedicated to giving the result and commenting on their properties.
The distribution of local time is determined in the
Section~\ref{distribution} using the 
technique presented in paragraph~\ref{maths2}.
The distribution in dimension one is given in~\ref{dist1}.
To obtain regular distributions in dimensions higher than one, 
it is necessary to introduce a small length scale
corresponding to the resolution. 
I compute the fluctuations of the local time in the other paragraph of
Section~\ref{distribution} 
and show that it depends on the resolution for dimensions higher than one.
The applications of these results in experiments are discussed
in Section~\ref{applications}.
Possible improvements are evoked in the conclusion.
The reader interested only in the results for
applications may focus on the text and 
the results of equations~\eqref{L1}, \eqref{L2}, \eqref{L3}, 
\eqref{fluct1d}, \eqref{fluct2d} and \eqref{fluct3d} 
and skip the technical paragraphs~\ref{GRqu}, \ref{maths1}, \ref{maths2}.

\section{Properties of the local time}
\label{general}
Let us consider a disordered medium in which particles or waves
travel following random trajectories. 
The distribution of probability for the particles' 
random trajectories coincides with the energy statistics of partial waves.
The particles or the waves are emitted from the origin at time~$t_0$
and are scattered by the heterogeneities of the medium.
The source term is therefore a Dirac delta function in space and time.
At time~$t$, the probability to find a particle 
-- or the fraction of wave energy -- in an elementary 
volume~$\ud[d]\vx$ centered at the position~$\vx$ is noted 
$G(\vx,\,t)\ud[d]\vx$. 
$G(\vx,\,t)$ is the elementary solution -- or Green's function -- 
of the \emph{transport equation}, a partial 
differential equation such as the diffusion
equation or the radiative transfer equation~\cite{chandrasekar}.

\begin{figure}
\begin{center}
\includegraphics[width=0.3\textwidth]{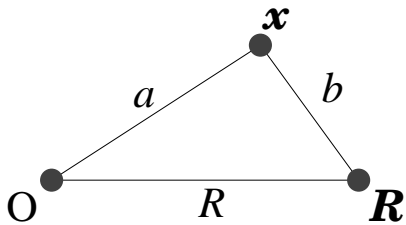}
\caption{\label{geometrie}Coordinates used in the article. $\vR$
is a fixed point at distance~$R=\|\vR\|$ from the origin. The 
generic point~$\vx$ is located at the distances~$a=\|\vx\|$ 
from the origin and $b=\|\vx-\vR\|$ from~$\vR$
respectively (dimension $d\geqslant2$). }
\end{center}
\end{figure}

A detection device at the position~$\vR$ is sensitive only to~$G(\vR,\,t)$
where~$\vR$ is in the close neighborhood of~$\vR$ such that at
a given time~$t$, it only detects the signal sent from the origin 
reaching~$\vR$ at this time. The signal recorded in~$\vR$
is thus the Green's function, or impulse response, from
the origin to $\vR$. During time~$t$, the observed signal is a superimposition
of signals due to random trajectories that have explored the medium.  As an
example, consider photons emitted in a pulse at time~$t=0$ 
from a source located at the origin. 
A photo-detector located in~$\vR$ measures the intensity $I(t)$ along time.
We suppose that the fraction of photons 
absorbed by the photo-detector represent a negligible fraction 
of the measured intensity and does not affect the experiment.
Suppose that after this measurement, the
absorption~$\mu(\vx)$ changes non-uniformly by the quantity~$\Delta\mu(\vx)$.
We now denote by~$I'(t)$ the intensity measured in~$\vR$ after these
changes if the source sends a rigorously identical pulse.
If photons spend a time~$\delta t$ at the position
$\vx$, their contribution to the logarithm of the intensity will 
be reduced by~$\Delta\mu(\vx)\delta t$. The average local 
time~$L(\vx,\,t)\ud[d]\vx$ is
the time spent on average by the photons in the elementary volume~$\ud[d]\vx$
around~$\vx$ during the whole propagation time~$t$ after the pulse emission. 
The logarithm of the intensity is therefore 
reduced by~$\Delta\mu(\vx)L(\vx,t)\ud[d]\vx$. 
As a conclusion the total observed logarithm of intensity will be
modified by
\begin{equation}
  \log\frac{I'(t)}{I(t)}=-\iiint\Delta\mu(\vx)L(\vx,t)\ud[d]\vx.
\label{intensity}
\end{equation}
The observed intensity is therefore modified by the average local time.
Supposing that we had not only a source and a receiver but many of each,
an imaging process would consist in measuring the variation of intensity
between each source and each receiver and performing a mathematical inversion
to reconstruct the value of~$\Delta\mu(\vx)$. 
In this article, I compute the expressions of~$L$ 
in diffusive media and its fluctuations. By symmetry considerations, 
it is easy to see that these quantities only depend on the 
distances~$R=\|\vR\|$, $a=\|\vx\|$ and $b=\|\vx-\vR\|$ at position~$\vx$ 
(see the figure\ref{geometrie}).

Let us now mathematically define the local time. 
Consider a function $f$ defined on~$\R^d$ and a random continuous
process~$\vX_u$ for~$0\leq u\leq t$, for instance
a propagation process. If the stochastic average
of~$\int_0^tf(\vX_u)\ud{u}$ over all realisations of~$\vX$ is expressed as
a \emph{space} integral of~$f$ weighted by a function~$L_d(\vx,\,t)$,
\begin{equation}
\moy{\int_0^tf(\vX_u)\ud{u}}=\int_{\R^d}f(\vx)L_d(\vx,\,t)\ud[d]\vx,
\label{def convol}
\end{equation}
then~$L_d(\vx,\,t)$ is called the \emph{average local time} of the
process~$\vX$. The brackets~$\moy{\cdot}$ denote
the stochastic average over the realisations of~$\vX$. 
Equation~\eqref{def convol} is sometimes called the occupation time 
formula~\cite{revuzyor}. 
Such a formula was originally imagined
by Trotter~\cite{trotter1958}, inspired by L\'evy's ``mesure de voisinage''
\cite{levy1965}. The usefulness of this formula is to transform a 
stochastic average into a regular integral. The average local time is
a positive, continuous function of space. Using the 
definition~\eqref{def convol} with the constant function~$f(\vx)=1$, we find
\begin{equation}
\int_{\R^d}L_d(\vx,\,t)\ud[d]\vx=t.
\label{fermeture}
\end{equation}
This relation expresses the intuitive fact that the particle spends in 
total the time~$t$ in the medium and that the average local density of 
this time is distributed according to the function~$\vx\mapsto L_d(\vx,\,t)$. 
In other words, Equation~\eqref{fermeture} translates into mathematics
the physical notion of average local time.
If we use now the definition~\eqref{def convol} with 
the function~$f(\vy)=\dirac[d]{\vx-\vy}$ we find the expression of the 
average local time as a stochastic average
\begin{equation}
L_d(\vx,\,t)=\moy{\int_0^t\dirac[d]{\vX_u-\vx}\ud u}.
\label{def L}
\end{equation}
The expressions~\eqref{def convol}, \eqref{fermeture} and~\eqref{def L}
do not depend on the stochastic process~$\vX_t$. They apply not only to
Brownian motion, but also for instance to the solutions of the radiative
transfer in a disordered medium~\cite{shang1988,sato1993,paasschens1997}. 
The local time of Brownian motion was studied mostly in Mathematics: 
The average local time has 
been explicitely computed in the case of a one-dimensional reflecting Brownian 
motion by Gittenberger and Louchard~\cite{gittenberger2000} 
and the probability distribution of 
local time in one dimension has been also determined by 
Pitman~\cite{pitman1999}. Other presentations of the local time of Brownian 
motion can be found in Ref.~\cite{revuzyor}.

\section{The Brownian bridge}
\label{bridge}
\subsection{Definition of the Brownian bridge}
\label{b1}
Let us consider a medium where the intensity is diffuse
$\derp tI(\vx,\,t)=D\nabla^2I(\vx,\,t)$. 
$D$ is the diffusion constant and fully characterizes the
transport in the medium. 

We consider a point source located at
the origin of the coordinate system. A receiver is located at point~$\vR$.
The distance of a point~$\vx$ to the origin and to~$\vR$ are denoted
respectively~$a$ and~$b$ (see Figure~\ref{geometrie}). 

In an infinite space of dimension~$d$
the diffusion kernel, also called elementary solution or Green's function,
at distance~$r$ and time~$t$
is given by the Gaussian distribution
\begin{equation}
 G_d(r,\,t)=\frac{1}{\left(4\pi D t\right)^{d/2}}
              \exp\left[-\frac{r^2}{4Dt}\right].
\label{diffusion}
\end{equation}
This distribution is equal to the probability distribution function
for the position of the point~$\vB_t$ of a Brownian motion starting
at the origin at time~$t=0$. 
For a given position~$\vR$ and a given time~$t$,
the \emph{Brownian bridge}~$\vP^{\vR,t}$ is
defined as the stochastic process
\begin{equation}
  \vP^{\vR,t}_u=\vB_u+\frac ut\left(\vR-\vB_t\right),
\label{def bridge}
\end{equation}
where~$0\leq u\leq t$ is the time variable.
The second term of the right-hand side of~\eqref{def bridge} ``forces'' 
the process to go to~$\vR$ at time~$t$.
The Brownian bridge~$\vP^{\vR,t}$ selects all the Brownian trajectories
satisfying the boundary conditions~$\vP^{\vR,t}_0=0$, $\vP^{\vR,t}_t=\vR$.
When computing the local time using the definition~\eqref{def L}, 
the statistical
average is performed over all the realizations of the Brownian bridge~%
$\vP^{\vR,t}$.
We prove this statement in the next, more technical paragraph,
along with other mathematical properties.

\subsection{Characteristic function of the drifted Brownian bridge}
\label{GRqu}
The Brownian bridge~\eqref{def bridge} is a Gaussian stochastic 
process of mean~$\frac ut\vR$.
The covariance of the Brownian motion~$\vB_u$ is 
$\moy{\vB_u\cdot\vB_v}=2D\min(u,v)$,
the covariance of the Brownian bridge with~$\vR=0$ is therefore
\begin{equation}
\moy{\vP^{0,t}_u\cdot\vP^{0,t}_v}=\frac {2D}tu(t-v),\quad(u\leq v). 
\label{covariance}
\end{equation}
As it is a Gaussian process, its characteristic function is
\begin{equation}
\phi_0(\vq,\,u)=\moy{\exp\ii\vq\cdot\vP^{0,t}_u}=
  \exp\left[-\frac Dtu(t-u)\vq^2\right].
\label{propag}
\end{equation}
The characteristic function for~$\vR\neq0$
is obtained from Eq.~\eqref{propag} using a drift
\begin{equation}
  \phi_{\vR}(\vq,\,u)=
    \exp\left[\ii \frac ut \vq\cdot\vR 
                 -\frac {u(t-u)}{t} D\vq^2\right].
  \label{G(q,u)}
\end{equation}

The usual properties of the characteristic function state that
for an arbitrary function~$f$, of~$d$-dimensional Fourier transform
$\tilde f$, we have
\begin{equation}
\moy{f(\vP^{\vR,t}_u)}=\int\frac{\ud[d]\vq}{(2\pi)^d}\,
  \phi_{\vR}(\vq,\,u)\;\tilde f(\vq).
  \label{GR}
\end{equation}

Summing the characteristic functions of~$\vP^{\vR,t}$ 
over~$\vR$ with the Gaussian weight~$G_d(\|\vR\|,\,t)$,
we obtain the characteristic function of the Brownian motion:
\[ \int_{\R^d}G_d(\|\vR\|,\,t)\,\phi_{\vR}(\vq,\,u)\ud[d]\vR
   =\tilde G_d(\vq,\,u), \]
which expresses the equivalence between Brownian motion and 
a collection of Brownian bridges. The Brownian bridge~$\vP^{\vR,t}$ 
selects the Brownian motion trajectories for which~$\vB_t=\vR$. 

\section{Average local time for diffusion}
\label{sec.average}

\subsection{General expression of the average local time}
\label{maths1}
The mathematics for the computation of the local time are detailed
in this paragraph. The results are given in the following paragraphs
of this Section.
I use as in Section~\ref{bridge} the notations~$a=\|\vx-A\|$, $b=\|\vx-B\|$ and
$R=\|B-A\|$. 
To compute the average local time~$L_d(\vx,t)$ from the 
equation~\eqref{def L} let us use the 
characteristic function formula~\eqref{GR} with 
the function~$f(\vy)=\dirac[d]{\vx-\vy}$: 
\begin{equation*}
  L_d(\vx,\,t)=\int_0^t \ud u \int_{\R^d}\frac{\ud[d]\vq}{(2\pi)^d}\,
     \phi_{\vR}(\vq,\,u) \;\e^{-\ii \vq\cdot\vx}.
\end{equation*}
The result is obtained by inserting the expression~\eqref{G(q,u)}
and performing the inverse Fourier transform. We get 
\begin{equation}
L_d(\vx,\,t)=\frac{t^{d/2} \e^{\frac{\vR^2}{4Dt}}}
  {\left(4\pi D\right)^{d/2}}
   \int_0^t
     \frac{\e^{-\frac{a^2}{4Du}}}{u^{d/2}}\;
     \frac{\e^{-\frac{b^2}{4D(t-u)}}}{(t-u)^{d/2}}\,\ud[]u.
\label{mu1}
\end{equation}
Let us relate this result to the solution of the diffusion
equation~\eqref{diffusion}
\begin{equation}
L_d(\vx,\,t)=\frac{1}{G_d(R,\,t)}\int_0^t\,G_d(a,\,u)G_d(b,t-u)\ud[]u.
\label{locadiff}
\end{equation}
The latter formula has a simple interpretation: the integrand is the
probability of going from~$A$ to~$\vx$ in time~$u$ and then from~$\vx$ 
to~$B$ in time~$t-u$. This probability is integrated over~$u$ since the
point~$\vx$ can be visited at any moment. The result is divided,
as required by the Bayes formula, by the probability of going
from~$A$ to~$B$ in time~$t$, the boundary conditions.
This formula is valid for any transport function~$G_d$ 
in multiple scattering media~\cite{rossetto2011}.
As diffusion is a Markov process, equality~\eqref{fermeture} follows
from the Chapman-Kolmogorov relation~\cite{gardiner} 
\begin{equation}
\int_{\R^d} G_d(a,\,u)G_d(b,\,t-u)\ud[d]\vx=G_d(R,t).
\label{transitivite}
\end{equation}

To compute the time convolution in Eq.~\eqref{locadiff}, I will
use the Laplace transform in time. Noting~$\hat f_j(s)$ the Laplace transform
of a function~$f_j(t)$, the Laplace transform of 
\[ \int_0^tf_1(u_1)\ud[]{u_1}\cdots \int_0^t f_n(u_n)
   \delta(t-u_1-\cdots-u_n)\ud[]{u_n} \]
as a function of~$t$ is equal to $\hat f_1(s)\cdots\hat f_n(s)$ 
(see Appendix~\ref{demo Laplace}). 
The Laplace transforms~$\hat G_d(a,\,s)$ 
for $d=1$, $2$ and $3$ are given in the 
table~\ref{Laplace123}.
In the Laplace domain, I include the 
weighting term~$G_d(R,\,t)^{-1}$ of~\eqref{locadiff} 
by defining
\begin{equation}
\Bar{f}(s)=\int_0^\infty \e^{-st}f(t)G_d(R,\,t)\ud[]t
\label{def Bar}
\end{equation}
to express Equation~\eqref{locadiff} undeg the simple form
\begin{equation}
\overline{L_d}(\vx,\,s)=\hat G_d(a,\,s)\hat G_d(b,\,s).
\label{locadiff L}
\end{equation}
The function~$\hat G_d(a,\,s)$ is singular in~$a=0$. The type of singularity
is denoted as~$g_d$ in the table~\ref{Laplace123} (see caption). 

\newcommand\ds\displaystyle
\begin{table}[h]
\begin{center}
\begin{tabular}{|c|c|p{4.5em}||c|}
\hline
$d$ & $\hat G_d(\|\vx\|,\,s)$ & Reference & Singularity $g_d(\vx)$ \\
\hline
$1$ & $\ds\frac{1}{\sqrt{4 D s}}
       \exp\left[-|x|\sqrt{\frac sD}\right]$ & 29.3.84
    & $\ds\frac12|x|$ \\
\hline
$2$ & $\ds \frac1{2\pi D}K_0\left(\|\vx\|\sqrt{\frac sD}\right)$ & 
  29.3.120 
  & $\ds\frac{1}{2\pi}\ln\|\vx\|$ \\
\hline
$3$ & $\ds \frac1{4\pi D \|\vx\|}\exp\left[-\|\vx\|\sqrt{\frac sD}\right]$ &
  29.3.82 
  & $\ds-\frac{1}{4\pi\,\|\vx\|}$ \\
\hline
\end{tabular}
\caption{\label{Laplace123}
Table of Laplace transforms in time of the diffusion 
kernel~\eqref{diffusion} for the dimensions~$1$, $2$ and~$3$. The numbers
refer to the equation in the handbook by Abramowitz and 
Stegun~\cite{abramowitzstegun}.
When~$\vx\to0$, $\hat G_d(\|\vx\|,\,s)$ is equivalent to
$g_d(\vx)/D$
where the singularity~$g_d(\vx)$ is the 
Green's function of the Laplace equation
in~$d$ dimensions. 
} 
\end{center}
\end{table}

\subsection{Average local time in one dimension}
To obtain the expression of the average local time
in one dimensions~($d=1$), we compute the
inverse Laplace transform of
\[ G_1(a,\,s)G_1(b,\,s)=\frac1{4Ds}\exp\left[-(a+b)\sqrt{\frac sD}\right], \]
which is, according to the reference~%
\cite[eq. 29.3.83]{abramowitzstegun}, a complementary error function.
Introducing the notation~$\erfc(x)=2/\sqrt{\pi}\int_x^\infty \exp[-u^2]\ud u$, 
we have
\begin{equation}
L_1(x,\,t)=\sqrt{\frac{\pi t}{4D}}
           \erfc\left[\frac{a+b}{\sqrt{4Dt}}\right]
           \exp\left[\frac{R^2}{4Dt}\right].
\label{L1}
\end{equation}
A particularity in expression~\eqref{L1} is that 
for all~$x$ between~$A$ and~$B$, $L_1$ has the same 
value because $a+b=R$ then~; 
This is a consequence of the Markovian character of the diffusion
kernel: as long as the process is between~$A$ and~$B$, it
has an equal probability to visit any position.
Let us also remark that the average local time~$L_1$ remains
finite in~$A$ and~$B$, but that is these points, there is
a slope change, as for $g_1(x)$.

When $A$ and $B$ coincide, we have~$R=0$ and $a=b$. In this case,
we note the local time~$\ell_d(a,\,t)$. In dimension
one we have therefore $\ell_1(a,\,t)=\sqrt{\pi t/4D}\erfc(a/\sqrt{Dt})$.

\subsection{Average local time in two dimensions}
\label{dim2}
To compute the exact result in two dimensions ($d=2$),
let us use the following result, demonstrated in the appendix~\ref{IK0}
\begin{equation}
 \int_0^1 \frac{\e^{-\alpha/u-\beta/(1-u)} \ud u}{u(1-u)}
    =2\e^{-\alpha-\beta}K_0\left(2\sqrt{\alpha\beta}\right). 
 \label{I0ab}
\end{equation}
$K_0$ is the modified Bessel function of the second kind and
of order~$0$.
Replacing~$\alpha$ by~$a^2/4Dt$ and~$\beta$ by~$b^2/4Dt$ and 
introducing the result of~\eqref{I0ab} in the Formula~\eqref{locadiff}
we obtain the local time kernel in dimension~2 
\begin{equation}
L_2(\vx,\,t)
  =\frac{1}{2\pi D}\exp\left[\frac{R^2-a^2-b^2}{4Dt}\right]
           K_0\left(\frac{ab}{2Dt}\right).
\label{L2}
\end{equation}
Remark that the average local time~$L_2$ has a logarithmic
divergence in~$A$ and~$B$, like~$g_2(\vx)$ at the origin.
When $A$ and $B$ coincide, the local time is expressed 
by~$\ell_2(a,t)=\frac{1}{2\pi D}
\exp\left[-\frac{a^2}{2Dt}\right]K_0\left(\frac{a^2}{2Dt}\right)$ as
already shown by Pacheco and Snieder \cite{pacheco2005}.

\subsection{Average local time in three dimensions}
In three dimensions ($d=3$), the same method as in one 
dimension is used. 
Remarkably the product
\begin{equation}
  \hat G_3(a,\,s)\hat G_3(b,\,s)=
     \frac{\exp\left[-(a+b)\sqrt{\frac sD}\right]}{(4\pi D)^2 \,ab}
\label{produit Lg3}
\end{equation}
is equal to~$\hat G_3(a+b,\,s)$
multiplied by a constant. The inversion is straightforward and yields
\begin{equation}
L_3(\vx,\,t)=\frac{a+b}{4\pi D \,ab}
             \exp\left[\frac{R^2-(a+b)^2}{4Dt}\right].
\label{L3}
\end{equation}
In this expression the time behaviour is exclusively contained in the
exponential. Note that the argument of the exponential is negative
thanks to the triangular inequality~$a+b\geqslant R$ for all~$\vx$.
Remark also that the average local time in three dimensions diverges
in~$A$ as~$a^{-1}$ and in~$B$ as~$b^{-1}$ like~$g_3(\vx)$ at the origin.
For coincident~$A$ and~$B$ we have 
$\ell_3(a,t)=\frac{1}{2\pi Da}\exp\left[-\frac{a^2}{Dt}\right]$
which was also given by~Pacheco and Snieder~\cite{pacheco2005}.

\section{Fluctuations of local time for diffusion}
\label{distribution}

\subsection{The origin of the fluctuations of local time}
If one considers single Brownian trajectories separately, the local time
in~$\vx$ depends on the particular realization of the Brownian motion.
Local time is therefore a random quantity and the origin of its
randomness is the different realizations of Brownian motion.
However, in a perfect diffusive media each
trajectory is realized by a partial wave which carries an energy proportional 
to its probability. In such a perfect medium, the local time would always be 
equal to its average value. 
Yet, actual diffusive media are not perfect since they have
a correlation length, called the \emph{mean free path}. As a consequence,
some Brownian trajectories are not followed by any partial waves. Different
realizations of the medium (sometimes referred to as realizations 
of the medium's disorder) select and exclude different Brownian trajectories.
Only a sample of all Brownian trajectories is realized in a given medium,
the value of the local time therefore differs from the results of
the previous Section and the difference depends on the positions
of all the constituents of the material.
Local time is therefore a random
variable in wave propagating media as well, but its randomness roots in
the different realizations of disorder. The sample space is made of
the configurations of the medium, and not of the realizations of a stochastic
process. Ergodicity states that these sample spaces lead to the same statistics.
In this section, I compute the probability distribution
function of the local time of a Brownian bridge and thanks to the ergodic
equivalence, use it to compute the fluctuations of the local time
with respect to the medium's disorder.

\subsection{Moments of the local time distribution}
\label{maths2}
In this paragraph, we perform the computation of the moments 
of the local time distribution. The results are
presented and discussed in the following, less technical, paragraphs. 
We start from the definition of the moment of order~$m$ of the 
local time distribution in dimension~$d$
\begin{equation}
  \mu_d^m(\vx,\,t)=\moy{\left(\int_0^t\,
     \delta^{(d)}\left(\vP^{\vR,t}_u-\vx\right) \ud u\right)^m}.  
\label{count1}
\end{equation}
When~$m=0$, the moment is~$\mu_d^0=1$. In the following developments,
we will assume $m\geq1$. 
Let us expand every Dirac $\delta$-function as a Fourier
integral
\begin{equation*}
\mu^m_d(\vx,\,t)=\int \prod_{j=1}^m \ud[]{u_j} \prod_{j=1}^m
  \frac{\ud[d]{\vq_j}}{(2\pi)^d}
  \moy{\e^{\ii\sum_{j=1}^m\vq_j\cdot(\vP^{\vR,t}_{u_j}-\vx)}}.
\end{equation*}
Let us reorder the times such 
that~$u_0=0\leq u_1\leq\dots\leq u_m\leq u_{m+1}=t$. 
The brackets in the above equation contain
the characteristic function of a Brownian bridge visiting~$\vx$ at times
$u_1\leq\,\dots,\leq u_m$. Therefore it is a sequence of Brownian 
bridges, the first between~$(A,\,u_0)$ and $(\vx,\,u_1)$,
then $m-1$ between~$(\vx,\,u_i)$ and~$(\vx,\,u_{i+1})$ for~$1\leq i\leq m-1$
and the last one between~$(\vx,\,u_m)$ and~$(B,\,u_{m+1})$. 
The characteristic function is thus Gaussian 
with correlation matrix~$C_{ij}=2Du_i(t-u_j)/t$, according to
the equation~\eqref{covariance}. The inverse Fourier transform leads to an
expression with the same structure as the Equation~\eqref{mu1},
noting~$a_1=a$,~$a_{m+1}=b$ and~$a_i=0$ for~$2\leq i\leq m$,
\begin{equation*}
\mu_d^m(\vx,\,t)=\frac{m!}{G_3(R,\,t)}
  \int\prod_{i=1}^m\ud[]{u_i} \prod_{i=1}^m G_d(a_i,\,u_{i+1}-u_i).
\end{equation*}
The integral is performed for the values of~$u_i$ obeying
$0=u_0\leq u_1\leq\cdots\leq u_{m+1}=t$. Let us change the variables~$u_i$
into~$v_i=u_{i+1}-u_i$ such that the condition on times is now~$\sum_0^mv_i=t$.
In the Appendix~\ref{demo Laplace}, I show that the Laplace transform
of this integral is a product of Laplace transforms
\begin{equation}
\overline{\mu_d^m}(\vx,\,s)=m!\,
\hat G_d(a,\,s)\hat G_d(b,\,s)\hat G_d(0,\,s)^{m-1}.
\label{Laplace mu}
\end{equation}
In the case~$m=1$ we retrieve the expression for the local 
time~\eqref{locadiff} ($\mu_d^1=L_d$).
The moments~$\overline{\mu_d^m}(\vx,\,s)$ depend on the
order~$m$ like $m! \alpha^m$, for some number~$\alpha$,
which is characteristic for an exponential distribution 
of width~$\alpha$. Denoting by~$\tau$ the value of the local time,
we deduce that its probability distribution is
\begin{align}
 \overline{p_d}(\vx,\,s;\,\tau)&=
  \overline{S_d}(\vx,\,s)\delta(\tau)+
 \frac{\overline{L_d}(\vx,\,s)}{\hat G_d(0,\,s)^2}
    \e^{-\frac{\tau}{\hat G_d(0,\,s)}}, 
\label{p(l)}\\
\overline{S_d}(\vx,\,s)&=\hat G_d(R,\,s)-\frac{\overline{L_d}(\vx,\,s)}
                        {\hat G_d(0,\,s)}.
\label{S}
\end{align}
Observe that the distribution of local time displays a singularity
at~$\tau=0$ which expresses the fact that some trajectories do
not visit the neighbourhood of~$\vx$. The term~$\overline S_d\delta(\tau)$
ensures the normalization of the probability distribution function.
I call this term the singular term and the second one the regular term
of the probability distribution function, denote it by~$\{p_d(\tau)\}$.

\subsection{Distribution of local time in one dimension}
\label{dist1}
In dimension~$d=1$, the expression~\eqref{p(l)} for~$\overline{p_1}$ 
gives for the regular part of the local time distribution
\begin{equation}
   \{\overline{p_1}(x,s;\,\tau)\}=
   \exp\left[-(a+b+2D\tau)\sqrt{\frac sD}\right]. 
\label{p1(l)}
\end{equation}
It is of the same form as~$\hat G_3$ so that the regular
local time probability distribution function in one dimension
\begin{equation}
 \{p_1(x,\,t;\,\tau)\}=
    \frac{a+b+2D\tau}t
    \exp\left(\frac{R^2-(a+b+2D\tau)^2}{4Dt}\right).
\label{p1}
\end{equation}
This result was first demonstrated by Pitman~\cite{pitman1999}.
The singular part is
\begin{equation}
\overline S_1(x,\,s)=\hat G_1(R,\,s)-
 \frac{\exp\left[-(a+b)\sqrt{\frac sD}\right]}{\sqrt{4Ds}}
\label{S1(s)}
\end{equation}
and is therefore a difference of two functions~$\hat G_1$. 
We obtain
\begin{equation}
S_1(x,\,t)=1-\exp\left(\frac{R^2-(a+b)^2}{4Dt}\right).
\label{S1(t)}
\end{equation}
$S_1(x,\,t)$ is the probability not to visit the position~$x$.
Observe that for~$x$ located between~$A$ and~$B$, this probability
is equal to~$0$ because one must visit it to cross from~$A$
to~$B$.

Writing~$\Phi(x)=\sqrt\pi\exp[x^2]\erfc(x)$ 
and~$y=(a+b)/\sqrt{4Dt}$,
we find the expression for the relative fluctuations of local time kernel 
in one dimension 
\begin{equation}
\frac{\Lambda_1(x,\,t)}{L_1(x,\,t)}=
  \left(\frac{4}{\Phi(y)^2}-\frac{4y}{\Phi(y)}-1\right)^{1/2}.
\label{fluct1d}
\end{equation}
The relative fluctuations as a function of~$y$ are rather large, 
even for long times,
the minimal value is~$\sqrt{4/\pi-1}\simeq0.52272$ when $y=0$. 

\subsection{Fluctuations of local time in two dimensions}
In two dimensions ($d=2$), the expression~\eqref{Laplace mu}
must be handled with care, because the expression of~$\hat G_2(a,\,s)$
has a logarithmic divergence when~$a\to0$. 
The inverse Laplace transform of the expression~\eqref{p(l)} 
does not exist. To regularize this divergence, let us introduce
a small length~$\varepsilon$ and replace~$\hat G_2(0,\,s)$ 
by~$\hat G_2(\varepsilon,\,s)$. 
The fluctuations are obtained from the integral 
\begin{equation}
  \int_0^t\ud[]u \int_0^{t-u}\ud[]v\;
  G_2(\varepsilon,\,u)\,G_2(a,\,v)\,G_2(b,\,t-u-v)
\end{equation}
that I compute in the Appendix~\ref{appendix mu2}.
The divergence in the small distance~$\varepsilon$ is
logarithmic and we obtain
\begin{equation}
\mu_2^2(\vx,\,t)\simeq-\frac{L_2(\vx,\,t)}{2\pi D}\ln\frac{\varepsilon^2}{4Dt}
\label{mu22}
\end{equation}
and the relative fluctuations are approximated by 
\begin{equation}
\frac{\Lambda_2(\vx,\,t)}{L_2(\vx,\,t)}\simeq
\left(\frac{-\ln\frac{\varepsilon^2}{4Dt}}{2\pi D \,L_2(\vx,\,t)}\right)^{1/2}.
\label{fluct2d}
\end{equation}
Let us remark that the fluctuations diverge for small resolution lengths
with the same behaviour as the square root of the average local time
for small distances from the source or the receiver.

The regularization length~$\varepsilon$ is necessary because in 
dimensions higher than two, 
Brownian motion visits a given point with probability~0.
This is observed in the fact that~$S\to1$ for~$\varepsilon\to0$. 
To ensure that the measured local time is not identically zero, it is necessary
to introduce~$\varepsilon$, the radius of a small ball centered
at~$\vx$ in which the local time is computed. Therefore, $\varepsilon$ is the
\emph{resolution} length. The average local time
remains finite in the limit~$\varepsilon\to0$ by construction,
but the moments of higher orders do not.

\subsection{Distribution of local time in three dimensions}
The local time distribution in three dimensions requires, 
as in the two dimensional case,
a regularization, because~$G_3(a,\,s)$ diverges when~$a\to0$. 
After regularization we obtain
$\mu_3^m\simeq L_3 \,m! (4\pi D\varepsilon)^{m-1}$ 
and therefore we get an exponential
distribution for all~$\varepsilon>0$. 
Writing~$\lambda=4\pi D \varepsilon$, we have
\begin{equation}
p_3(\vx,\,t;\,\tau)=(1-\lambda\;L_3)\delta(\tau)
  +\lambda^2\,L_3\,\exp(-\lambda\tau).
\label{p3}
\end{equation}

From the equation~\eqref{p3} we obtain the fluctuations of the
measured local time in three dimensions. The result
\begin{equation}
\frac{\Lambda_3(\vx,\,t)}{L_3(\vx,\,t)}=\left(\frac{1} 
   {2\pi D \varepsilon\; L_3(\vx,\,t)}-1\right)^{1/2}
\label{fluct3d}
\end{equation}
states that the fluctuations of the local time in three dimensions are large
and grows as the resolution length decreases.
Notice that, as in two dimensions, the fluctuations diverge for
small resolution lengths with the same behaviour as the square root 
of the average local time close to the source or the receiver.

\section{Applications of local time to imaging}
\label{applications}
\begin{figure}
\includegraphics[width=0.45\textwidth]{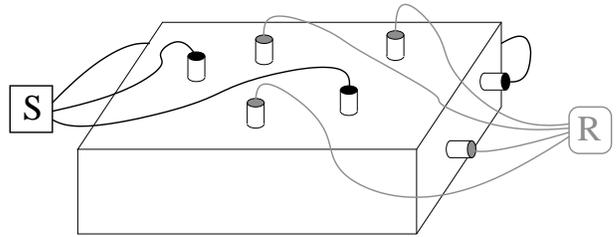}
\caption{\label{setup}Sketch of an experimental setup with three sources
(S, black) and four receivers (R, grey) creating $n=12$ pairs. 
The source and the receiver must be synchronized. The receivers can be
used simultaneously, but the sources should be activated one at a time.
If the devices of a pair
are too distant, there may be no usable signal. In such case, the pair can 
be removed from the input vector~$M$ and the kernel~$K$. }
\end{figure}

In disordered media, the average local
time is a measure of the sensitivity of the wave field to
a change. The average local time therefore provides the solution
to the \emph{direct problem}: How do the changes in the medium
modify the wave fields ?
In imaging, however, the changes are unknown and the waves are sent
into the medium to monitor them. Thanks to 
the expressions of the average local time~\eqref{L1}, \eqref{L2}
and~\eqref{L3}, one can retrieve informations about the changes
from the measurement of wave fields of intensity: 
this is the \emph{inverse problem}.
To perform such inversions, it is necessary to obtain several
independant measurements of the systems by installing
several sources or receivers. A sketch of an experimental setup is
shown in Figure~\ref{setup}.
Each pair formed by a source and a
receiver provides a measurement of an average local quantity weighted
by the local time. Because local times for each pair are different
at a given point the inversion is possible. The images produced
by such inversions are maps in one, two or three dimensions of the
medium.  Such inversions have been already performed to image the changes of
velocity (CWI)~\cite{obermann2013}
or scattering (LOCADIFF)~\cite{larose2010}. 
Note that only the intensity is required to produce an image of the 
absorption variation (AV), in which case the average local time 
imaging technique extends to optics.

In disordered media, the detection and location
of strong changes is not a challenge any more: using large wavelength
allows using standard imaging techniques. The local time is therefore
useful in the case where the changes occurring in the medium are of
small amplitude. In such situations, the effects of two separate changes
add up, as it was shown for AV in the introduction 
because the expression~\eqref{intensity} is linear
in~$\Delta\mu(\vx)$. In the case of CWI, time delays
add up as long as~$\delta v/c \ll 1$, where $\delta v$ is the
velocity change and $c$ the average velocity.
In LOCADIFF, the losses of correlation due to distant scatterers
add up as long as~$c\Delta\sigma L(\vx,t)\ll1$, where $\Delta\sigma$
is the change of scattering cross-section~\cite{rossetto2011}.
In these three situations, one obtains a linear relation such as 
Equation~\eqref{intensity}. 

The linear relation is the same in the case of AV, CWI and LOCADIFF.
The nature of the input and output physical quantities 
in different techniques is sum up in the Table~\ref{techniques}.
In a practice the medium is divided into a large 
number~$N$ of elements, called \emph{voxels}, 
centered at~$\vx_j$ ($1\leq j\leq N$) of volume~$\delta V_j$.
The medium is equipped with~$n$ source-receiver pairs.
I denote by~$\vS_i$ and~$\R_i$ the positions
of the source and the receiver of the pair~$i$ respectively. 
The ``input'' vector~$M$ is made of~$n$ independent measurements
and the output image is a vector~$Q$ of size~$N$.
The relation between the measurements~$M$ and the image~$Q$ is
a matrix equation
\begin{equation}
KQ=M
\label{imaging}
\end{equation}
in which~$K$ is a~$n\times N$ matrix called the~\emph{kernel}.
The element~$K_{ij}$ of the kernel is obtained using
$a=\|\vx_j-\vS_i\|$, $b=\|\vx_j-\R_i\|$ and $R=\|\R_i-\vS_i\|$ in 
the appropriate average local time formula:
$K_{ij}=L(\vS_i,\,\vx_j,\,\R_i\,,t)\delta V_j$.
In principle, one should install as many source-receiver
pairs as the number of voxels of the images to ensure that
equation~\eqref{imaging} has a solution.
 However, inversion techniques such
as the error-minimization algorithm of~Tarantola-Valette~\cite{tarantola1982},
linear minimization or compressive sensing techniques~\cite{candes2006}
provide in general satisfying results from incomplete informations
when $n\ll N$. Using extra informations, such as for instance~$Q\geq 0$ in 
the case of LOCADIFF, allows to improve the efficiency and the accuracy of
the inversion. It also proves useful to assume \emph{sparsity},
that is to say that the changes are
localized, such that~$Q=0$ on a large majority of voxels.
The assumption of sparsity increases the ratio $N/n$. 

\begin{table}
\begin{tabular}{|c|c|c|p{0.23\linewidth}|}
\hline
Technique & Input ($M$) & Images ($Q$) & \hfil Waves\\
\hline
\hline
CWI & relative delay  & relative velocity & sound, elastic \\
\hline
LOCADIFF & correlation loss & scattering & sound, elastic\\
\hline
AV & log (intensity) & absorption & sound, elastic, light \\
\hline
\end{tabular}
\caption{\label{techniques}Different 
imaging techniques using the average local
time kernel. ``Input'' is the nature of the measurement,
``Images'' refers to the quantity one creates the image of.
The column ``Waves'' indicates to which waves the technique
can be applied. For elastic waves, this holds when waves are
in the equipartition regime, which is a consequence of diffusion.
The coda wave interferometry technique (CWI)
uses time stretches on the wave field records to measure small
velocity changes~\cite{pacheco2005}. 
The LOCADIFF technique uses the correlation losses
and produces images of the changes of scattering cross-section,
that is to say the structural changes~\cite{larose2010,rossetto2011}.
The absorption variation (AV) has been described in Section~\ref{general}.
Note that as it is based on the intensity variations, the expected sensitivity
of AV is smaller than for the other techniques.}
\end{table}

The reliability of the images produced
by inversion is in most cases extremely important, as is their
accuracy too. In two and three dimensions, the most common experimental
cases, the resolution length~$\varepsilon$ has to be chosen for an optimal
quality of imaging balancing a low resolution for large values of
$\varepsilon$ and inaccurate results for small values of~$\varepsilon$. 
In the algorithm
of Tarantola-Valette, the errors on the input measurements enter
as a parameter of the inversion. The inversion provides an estimate
of the errors on the results. One can therefore obtain, using the
expressions for the fluctuations computed in the section~\ref{distribution},
tune the resolution length for optimal results.
The fluctuations -- which expressions are provided 
by Equations~\eqref{fluct1d}, \eqref{fluct2d} and~\eqref{fluct3d} --
are due to the randomness of the medium and can therefore not be disposed
of by repeating the measurements. 
To obtain independent, numerous enough
input data, it is necessary to increase the number of measurement devices
or use several disjoint time windows in the coda. The latter solution
is however restricted by the quality of the recordings and the duration
of the coda. As a conclusion from these remarks, it appears that the inverse
problem based on a local time kernel is delicate and requires a good
understanding of the different aspects of the problem.

\section{Conclusion}
\label{conclusion}
In this article, I have introduced an elementary theory of local
time for transport and its statistics. The average local time in 
infinite diffusive media and its fluctuations have been computed exactly
for the dimensions one, two and three with a technique that is easily
extended to higher dimensions in the appendix. 
Beyond the explicit expressions, it is important to note that
the fluctuations explicitely depend on the resolution length, and diverge
as the square root of the Green's function of the Laplace equation.

Wave propagation has a finite velocity which is not taken into 
account in the diffusion approximation. A more rigorous transport
equation than the diffusion equation is the \emph{radiative transfer equation}
\cite{chandrasekar}. Solutions to this equation
are known in dimensions one, two and three~\cite{paasschens1997}
and differ from the diffusion only at short times. 
Clipping the integration domain of~\eqref{locadiff} is not correct
because it does not account for the conservation of energy. The
solutions to the radiative transfer contain ballistic
terms that compensate this energy loss and preserve the conservation
of energy.
The short time discrepancy between diffusion and radiative transfer
will change the average local time and its distribution close
to the source and the receiver at distances of the order of the
mean free path.
A future improvement of this work is
therefore turned to the radiative transfer solution of the wave
equation in disordered medium. The solution to the two-dimensional
radiative transfer equation has been already used to numerically compute 
the local time~\cite{obermann2013}. 

Even though diffusion screens
the medium's boundary conditions beyond a few mean free paths, these
conditions are important in situations where the changes occur close
to them. If the boundary is absorbing or reflecting, the average local time 
will
be respectively reduced or increased. Simple boundary conditions, 
where the method of images applies, have been extensively discussed 
in Ref.~\cite{rossetto2011}. 

Promising results \cite{larose2010,obermann2013,obermann}, however, show that
applications of local time to imaging offers new insights in physical 
sciences of complex media.

\acknowledgements
This work was funded by grant number JC08-313906 from ANR
and SMINGUE-UJF.
The author whishes to thank {\'E}ric Larose and Thomas Plan{\`e}s
for stimulating discussions.

\appendix
\section{Laplace transform of time convolutions}
\label{demo Laplace}
I note the multiple convolution of~$n$ functions~$f_j$
\begin{equation*}
h(t)=\int_0^tf_1(v_1)\ud[]{v_1}\cdots \int_0^t f_n(v_n)
   \delta(t-v_1-\cdots-v_n)\ud[]{v_n}.
\end{equation*}
The integral can be performed from zero to infinity, since
the Dirac distribution ensures that only values of~$v_j$
between~$0$ and~$t$ contribute to the result. 
The Laplace transform of the function~$t\mapsto\delta(t-v_1-\cdots-v_n)$
is~$\exp(-sv_1-\cdots-sv_n)$
therefore the Laplace transform of~$h(t)$ is 
\begin{equation*}
\hat h(s)=\int_0^\infty f(v_1)\ud[]{v_1}\cdots\int_0^\infty f(v_n)\ud[]{v_n}
     \e^{-sv_1-\cdots-sv_n}.
\end{equation*}
The integrals are independent and each of them yields the
Laplace transform of a function~$f_j$. Therefore
\begin{equation}
\hat h(s)=\hat f_1(s)\cdots\hat f_n(s).
\label{produit L}
\end{equation}

\section{Computation of $I_n(x,y)$}
\label{IK0}
Let us compute the integrals
\[ I_n(\alpha,\,\beta)=\int_0^1 \e^{-\alpha/u} \e^{-\beta/(1-u)} 
  \frac{\ud u}{u^{n+1}{(1-u)}^{n+1}} \]
using the new variable~$x=\frac{1-u}u$. We obtain
\[ I_n(\alpha,\,\beta)=\e^{-\alpha-\beta}\int_0^\infty
  \frac{(x+1)^{2n}}{x^{n+1}}\; \e^{-\alpha x-\frac\beta x}\, \ud x.\]
I note~$J_p(\alpha,\,\beta)=\int_0^\infty
    \exp(-\alpha x-\beta/x)x^{-p-1}\ud x$.
The value of~$J_p$ is found in the Gradsteyn \& Ryzhik 
table~\cite[integral 3.471.9]{gradsteynryzhik}~:
\begin{equation}
J_{-p}(\beta,\,\alpha)=J_p(\alpha,\,\beta)=
  2\left(\frac\alpha\beta\right)^{p/2}K_p\left(2\sqrt{\alpha\beta}\right),
\label{JabK}
\end{equation}
and we can express~$I_n(\alpha,\,\beta)$ for $n>0$ as a sum
\begin{multline}
 I_n(\alpha,\,\beta)=\binom{2n}{n}I_0(\alpha,\,\beta)+\\
  2\e^{-\alpha-\beta}\sum_{p=1}^{n}\binom{2n}{n-p}
\frac{\alpha^p+\beta^p}{(\alpha\beta)^{p/2}}K_p(2\sqrt{\alpha\beta}).
\label{Inab}
\end{multline}

\section{Computation of~$\mu_2^2$}
\label{appendix mu2}
The computation of~$\mu_2^2(\vx,\,t)$ is based on the integral
\[ M(\alpha,\,\beta,\,\epsilon)=
\int_0^1\ud[]u \int_0^{1-u}\ud[]v
  \frac{\exp\left[-\frac{\epsilon}{u}-\frac{\alpha}{v}-\frac{\beta}{1-u-v}
  \right]}{uv(1-u-v)} \]
We integrate~$M$ thanks to the change of variable~$x=u/(1-u)$ after integrating
over~$v$ as in Appendix~\ref{IK0}:
\[ M=2\e^{-\epsilon-\alpha-\beta}
 \int_0^\infty \e^{-(\alpha+\beta)x-\epsilon/x}
    K_0(2\sqrt{\alpha\beta}(1+x)) \frac{\ud[]x}x.\]
We use the Taylor expansion of~$K_0$
\[ K_0(2a(1+x))=\sum_{n,p\geq0}\binom np \frac{(-a)^nx^{n-1}}{n!}K_{n-2p}(2a) \]
in the integral. Performing the integration over~$x$ we
get terms~$J_{-n}(\alpha+\beta,\,\epsilon)$
defined by formula~\eqref{JabK}. For~$n\neq0$ and
$\epsilon\to0$, we have
$J_{-n}(\alpha+\beta,\,\epsilon)\simeq 2(n-1)! (\alpha+\beta)^{-n}$.
The series rewrite 
\begin{multline*}
   M\simeq4\e^{-\alpha-\beta}\left\{\vphantom{\sum_k^\infty}
             K_0(2\sqrt{\alpha\beta})
             K_0(2\sqrt{\epsilon(\alpha+\beta)})+\cdots\right.\\
   +\left.\sum_{n\geq1,p\geq0}\binom {n-1}p 
   \left(-\frac{\sqrt{\alpha\beta}}{\alpha+\beta}\right)^n 
    K_{n-2p}(2\sqrt{\alpha\beta})\right\}.
\end{multline*}
The leading term when~$\epsilon\to0$ is therefore logarithmic
\begin{equation}
M(\alpha,\,\beta,\,\epsilon) \underset{\epsilon\to0}\simeq
   -2\e^{-\alpha-\beta}K_0(2\sqrt{\alpha\beta})\,\ln\epsilon.
\label{equiv M}
\end{equation}

\section{Results in higher dimensions}
\label{higher}
Although for direct applications in physics the higher dimensions are of little
interest, some processes may be well described by a random walk in
high dimensions. In this appendix, I give less detailed results concerning
local times in dimensions higher than three. 
\subsection{Average local time in higher even dimensions}
\label{dim2n}

In a space of even dimension~$d=2n+2$ ($n>0$), we
compute the time convolution~\eqref{locadiff} using the same technique as 
for~$d=2$. We use the
expression~\eqref{Inab} of~$I_n(\alpha,\,\beta)$ and obtain
\begin{multline}
L_{2n+2}(\vx,\,t)=\frac{t}{(4\pi Dt)^{n+1}}
\exp\left[\frac{R^2-a^2-b^2}{4Dt}\right]\times\cdots\\
   \cdots\sum_{p=0}^n (2-\delta_p) \binom{2n}{n-p}
   \left(\frac{a^p}{b^p}+\frac{b^p}{a^p}\right)
   K_p\left(\frac{ab}{2Dt}\right)
\label{f2n+2}
\end{multline}
where~$\delta_p=1$ if~$p=0$ and $\delta_p=0$ otherwise.

\subsection{Average local time in higher odd dimensions}
\label{dim2n+1}
For the odd dimensions~$d=2n+3$, the we use the Laplace transform
of~$G_d$ in which the modified Bessel functions of the second kind are 
of half integral order and therefore have a
simplified expression~\cite[eq. 9.7.2]{abramowitzstegun}. 
We obtain 
\begin{equation}
\hat G_{2n+3}(a,\,s)=\pP_n\left(a\sqrt{\frac sD}\right)
\frac{\exp\left[-a\sqrt{\frac sD}\right]}{2(2\pi)^{n+1}Da^{2n+1}},
\label{Lg2n+1}
\end{equation}
where~$\pP_n$ is a unitary polynomial of degree~$n$
\begin{equation}
\pP_n(x)=\sum_{p=0}^n\frac{(n+p)!}{2^pp!(n-p)!}\;x^{n-p}.
\label{def pn}
\end{equation}

The product~$\hat G_{2n+3}(a,\,s)\hat G_{2n+3}(b,\,s)$
contains therefore the product of two polynomials~$\pP_n$, 
which is a unitary polynomial of degree~$2n$. 
Let us express it in the basis of the~$\pP_k$ polynomials in
$(a+b)x$:
\begin{multline}
\pP_n\left(a x\right)
   \pP_n\left(b x\right)=\\
   \frac{a^nb^n}{(a+b)^{2n}}
   \sum_{k=0}^{n} \pQ^n_k\left(\frac ab+\frac ba\right)
   \pP_{2n-k}\left((a+b)x\right),
\label{decomposition}
\end{multline}
where~$\pQ^n_k$ is a polynomial of degree~$\leqslant k$
(see table~\ref{qnk}).
Although a
general decomposition formula on this basis would require to sum up to~$k=2n$,
the sum is limited to~$n$ because the term of order~$k$ 
gives after Laplace inversion a terms
$G_{4n-2k+3}(a+b,\,t)/G_{2n+3}(R,\,t)$ which decreases as~$t^{k-n}$. 
For~$t\to\infty$, the local time must remain finite and therefore
one has~$k\leq n$. 

\begin{table*}
\begin{tabular}{|c||c|c|c|c|c|c|c|c|}
\hline 
$k$ & 0 & 1 & 2 & 3 & 4 \\
\hline
$\pQ^1_k(X)$ &1& $X-1$ & & &\\
$\pQ^2_k(X)$ &1& $3X-4$ & $3(X^2-X-1)$ & &\\
$\pQ^3_k(X)$ &1& $6X-9$ & $3(5X^2-8X-1)$ & $15(X^3-X^2-2X+1)$ & \\
$\pQ^4_k(X)$ &1& $10X-16$ & $9(5X^2-10X+2)$ & $15(7X^3-12X^2-6X+8)$ 
 & $105(X-1)(X^3-3X-1)$ \\  
\hline
\end{tabular}
\caption{\label{qnk}The $\pQ^n_k$ polynomials for $n\leqslant 4$. 
The value of $\pQ^n_0$ is always equal to~$1$.}
\end{table*}

The average local time kernel in arbitrary odd dimensions is therefore
given by
\begin{multline}
  L_{2n+3}(\vx,\,t)=\frac{a+b}{2D\,(2\pi ab)^{n+1}}
   \exp\left[\frac{R^2-(a+b)^2}{4Dt}\right]\cdots\\
  \cdots\times\sum_{k=0}^{n} \left(\frac{(a+b)^2}{2Dt}\right)^{n-k} 
          \pQ^n_{k}\left(\frac ab+\frac ba\right).
\label{L2n+3}
\end{multline}

\subsection{Fluctuations of local time in higher dimensions}
In higher dimensions as in dimensions~1, 2 and~3, the moment of
order~2 is proportionnal to the singularity~$g_d(\varepsilon)=
1/{\cal S}_d\varepsilon^{d-2}$ (${\cal S}_d$ is the surface of the
$d$-dimensional sphere). 

\begin{equation}
\frac{\Lambda_d(\vx,\,t)}{L_d(\vx,\,t)}\simeq\frac{1} 
   {\sqrt{D{\cal S}_d \,L_d(\vx,\,t)\; \varepsilon^{d-2}/2}}
\end{equation}

\end{document}